\documentclass[twocolumn]{aastex63}

\usepackage[fleqn]{amsmath}
\usepackage[hyphenbreaks]{breakurl}
\usepackage{float}
\usepackage{url}

\makeatletter
\newcommand*{\rom}[1]{\expandafter\@slowromancap\romannumeral #1@}
\makeatother

\submitjournal{AJ}

\shorttitle{Stellar distance indicators in the Local group galaxies}
\shortauthors{Bhardwaj A. et al.}
\begin{document}
\title{Stellar Variability and Distance Indicators in the Near-infrared in Nearby Galaxies. \\ I. RR Lyrae and Anomalous Cepheids in Draco dwarf spheroidal}

\correspondingauthor{Anupam Bhardwaj}
\email{anupam.bhardwaj@iucaa.in}
\author[0000-0001-6147-3360]{Anupam Bhardwaj}%\thanks{Marie-Curie Fellow}
\affil{Inter-University Center for Astronomy and Astrophysics (IUCAA), Post Bag 4, Ganeshkhind, Pune 411 007, India}
\author[0000-0002-6577-2787]{Marina Rejkuba}
\affiliation{European Southern Observatory, Karl-Schwarzschild-Stra\ss e 2, 85748, Garching, Germany}
\author[0000-0001-8771-7554]{Chow-Choong Ngeow}
\affil{Graduate Institute of Astronomy, National Central University, 300 Jhongda Road, 32001 Jhongli, Taiwan}
\author[0000-0002-1330-2927]{Marcella Marconi}
\affil{INAF-Osservatorio Astronomico di Capodimonte, Via Moiariello 16, 80131 Napoli, Italy}
\author{Vincenzo Ripepi}
\affil{INAF-Osservatorio Astronomico di Capodimonte, Via Moiariello 16, 80131 Napoli, Italy}
\author{Abhinna Sundar Samantaray}
\affil{Astronomisches Rechen-Institut, Zentrum fur Astronomie, Universit\"at Heidelberg, M\"onchhofstrasse 12-14, 69120 Heidelberg, Germany}
\author[0000-0001-6802-6539]{Harinder P. Singh}
\affiliation{Department of Physics and Astrophysics, University of Delhi,  Delhi-110007, India }
%\author{Shashi M. Kanbur}
%\affiliation{Department of Physics, State University of New York, Oswego, NY 13126, USA}
%\collaboration{(AAS Journals Data Scientists collaboration)}

\begin{abstract}

Draco dwarf Spheroidal galaxy (dSph) is one of the nearest and the most dark matter dominated satellites of the Milky Way. We obtained multi-epoch near-infrared (NIR, $JHK_s$) observations of the central region of Draco dSph covering a sky area of $\sim 21'\times21'$ using the WIRCam instrument at the 3.6-m Canada–France–Hawaii Telescope. Homogeneous $JHK_s$ time-series photometry for 212 RR Lyrae (173 fundamental-mode, 24 first-overtone, and 15 mixed-mode variables) and 5 Anomalous Cepheids in Draco dSph is presented and used to derive their period–luminosity relations at NIR wavelengths for the first-time. The small scatter of $\sim 0.05$~mag in these empirical relations for RR Lyrae stars is consistent with those in globular clusters and suggests a very small metallicity spread, up to $\sim0.2$~dex, among these centrally located variables. Based on empirically calibrated NIR period–luminosity–metallicity relations for RR Lyrae in globular clusters, we determined a distance modulus to Draco dSph of $\mu_\textrm{RRL} = 19.557 \pm 0.026$ mag. The calibrated $K_s$-band period-luminosity relations for Anomalous Cepheids in the Draco dSph and the Large Magellanic Cloud exhibit statistically consistent slopes but systematically different zero-points, hinting at possible metallicity dependence of $\sim-0.3$ mag~dex$^{-1}$. Finally, the apparent magnitudes of the tip of the red giant branch in $I$ and $J$ bands also agree well with their absolute calibrations with the adopted RR Lyrae distance to Draco. Our recommended $\sim1.5\%$ precise RR Lyrae distance, $D_\textrm{Draco} = 81.55 \pm 0.98 \textrm{(statistical)} \pm 1.17 \textrm{(systematic)}$~kpc, is the most accurate and precise distance to Draco dSph galaxy. \\
\end{abstract}

\section{Introduction}
\label{sec:intro}

Dwarf galaxies are the most numerous in the Universe and among the best astrophysical laboratories to study the structure of dark matter halos and the nature of dark matter \citep{bullock2017}. According to the Lambda cold dark matter model, smaller galaxies form first in the large-scale structure formation making dwarf galaxies ideal targets to constrain galaxy formation and evolution models \citep[see reviews by][]{tolstoy2009, battaglia2022}. Dwarf spheroidal (dSph) galaxies are low-luminosity ($M_V \geq -14$~mag), low-mass ($\sim10^7M_\odot$) early-type dwarf galaxies that host dominant old and intermediate-age stellar populations \citep{grebel2003, tolstoy2009}. The dSph galaxies with high mass-to-light ratios are particularly interesting systems to search for dark matter that can be inferred from analyses of their internal stellar kinematics \citep{mateo1998, mcconnachie2012}.

The nearest Galactic dSph galaxies ($D \lesssim 150$~kpc) are also ideal systems for detailed studies of individual stellar kinematics and their chemical abundances down to faint main sequence evolutionary phases. \citet{aaronson1983} measured stellar velocity dispersions of 4 Carbon stars in Draco dSph, which were found to be significantly larger than those of globular clusters. The author inferred the mass of the Draco to be much larger than its visible content providing the first hint that dSph galaxies contain a significant amount of dark matter. Since then stellar kinematics of hundreds of individual stars including their three-dimensional internal motions, have established Draco dSph as one of the most dark matter dominated satellites of the Milky Way \citep[e.g.,][]{read2018, massari2020, battaglia2022}.

Distance of a galaxy is the fundamental ingredient for modeling of its dynamical history, formation and evolution. Nevertheless, despite extensive studies of resolved stellar populations in Draco 
\citep[e.g.,][]{grillmair1998, aparicio2001, cioni2005, segall2007, kinemuchi2008, walker2015, muraveva2020} the distance estimates to this dSph available in the literature have a considerable scatter, which makes it poorly constrained. The distance modulus to Draco dSph ranges between $19.35$ and $19.85$ mag with typical uncertainties of $5-7\%$ (see Section~\ref{sec:dis}). The majority of previous distance determinations are based on horizontal branch stars and RR Lyrae (RRL) visual magnitude–metallicity  ($M_V -$ [Fe/H]) relations \citep{stetson1979,bellazzini2002,kinemuchi2008,muraveva2020}. However, these empirical relations may not be linear over the entire metallicity range, and are very sensitive to evolutionary effects because evolved RRL stars exhibit higher luminosities \citep{caputo2000, beaton2018, bhardwaj2020}. Furthermore, significant uncertainties in metallicity and reddening can lead to systematic errors of the order of $5\%$ in distance measurements based on optical data for RRL stars. 

High-precision distances based on RRL and Anomalous Cepheid  (ACEP) variable stars can be obtained at near-infrared (NIR) wavelengths where these pulsating stars exhibit tight period-luminosity (PL) and period-Wesenheit (PW) relations \citep[][and references therein]{longmore1986, ripepi2014a, bhardwaj2022, monelli2022}. At optical bands, variable star search in Draco dSph dates back to \citet{baade1961}, who discovered more than 260 variables including 133 RRL stars, which were used in several subsequent studies to improve their identification and classification \citep{zinn1976,  nemec1985b}. Modern $VI$ band photometric observations of variable stars in Draco dSph were provided by \citet{kinemuchi2008} including 270 RRL and 9 ACEP stars. \citet{muraveva2020} revisited RRL population in Draco dSph using {\it Gaia} data investigating their $M_G-$[Fe/H] relations. However, time-series observations for variable stars in Draco dSph have not been carried out so far at infrared wavelengths. We note that random-epoch NIR photometry of variable stars in several other Local Group galaxies have been utilized by the Araucaria project to determine their precise distances \citep{gieren2005, karczmarek2021}.

This paper is the first in a series of Stellar VAriability and Distance Indicators in the Near-infrared (SVADhIN\footnote{SVADhIN refers to freedom or independence in Hindi and Sanskrit languages.}) project, which aims to obtain NIR multi-epoch observations of variable stars in nearby Local Group galaxies. Homogeneous time-series NIR photometry of RRL and ACEP variables in the pilot Draco dSph will enable determination of a precise distance to this galaxy. The observations and data reduction are presented in Section~\ref{sec:data}. Light curves and PL relations for variable stars are discussed in Sections~\ref{sec:var} and \ref{sec:plrs}. The distance to Draco dSph is presented in Section~\ref{sec:dis} and the results are summarized in Section~\ref{sec:discuss}.

\section{Data and photometry} \label{sec:data}

\subsection{Observations and data reduction}

Multi-epoch NIR observations of Draco dSph were obtained using the WIRCam instrument \citep{puget2004} mounted on the 3.6-m Canada-France-Hawaii Telescope (CFHT). The NIR camera, WIRCam, is an array of four $2048\times2048$ HgCdTe HAWAII-RG2 detectors that are arranged in a $2\times2$ grid with gaps of $45\arcsec$ between adjacent detectors. The pixel scale is $0.3\arcsec$ pixel$^{-1}$ resulting in a wide field of view of $\sim21'\times 21'$. The multi-epoch $JHK_s$ observations were collected in the queue mode on 6 different nights between Feb 15, 2022 and Aug 8, 2022. We had requested 10 $JK_s$ epochs and 3 $H$ band epochs, but finally received 10, 5, and 8 epochs in $J$, $H$, $K_s$, respectively, due to a combination of the target's visibility, weather constraints and priority among allocated projects during the scheduled WIRCam runs. Each $J$ band epoch consisted of 20 dithered images with 30s exposure times, while 40 ditherered images were taken in $HK_s$ each with 25s exposure time.  A summary of all the epochs in $JHK_s$-bands is listed in Table~\ref{tbl:data}.  

\begin{deluxetable}{cccccc}
\tablecaption{Log of NIR observations. \label{tbl:data}}
\tablewidth{0pt}
\tablehead{\colhead{Date}  & \colhead{MJD} & \colhead{FWHM}  & \colhead{AM} & \colhead{ET}  & \colhead{$N_\textrm{f}$} }
\startdata
\multicolumn{6}{c}{$J$-band}   \\
\hline
         2022-02-15&    59625.6707&     1.73&     1.38&  30& 20\\
         2022-07-04&    59764.4427&     1.54&     1.36&  30& 20\\
         2022-07-04&    59764.5053&     1.85&     1.62&  30& 20\\
         2022-07-05&    59765.2795&     1.69&     1.41&  30& 21\\
         2022-07-05&    59765.4126&     1.65&     1.30&  30& 20\\
         2022-07-05&    59765.4818&     1.76&     1.51&  30& 20\\
         2022-07-06&    59766.3045&     1.35&     1.33&  30& 20\\
         2022-07-06&    59766.4590&     1.83&     1.43&  30& 18\\
         2022-08-07&    59798.3306&     1.46&     1.32&  30& 20\\
         2022-08-08&    59799.2769&     1.40&     1.27&  30& 20\\
\hline
\multicolumn{6}{c}{$H$-band}   \\
\hline
         2022-07-04&    59764.4736&     1.56&     1.46&  25& 40\\
         2022-07-05&    59765.4449&     1.67&     1.37&  25& 40\\
         2022-07-06&    59766.2576&     1.80&     1.48&  25& 28\\
         2022-07-06&    59766.4724&     2.00&     1.48&  25& 40\\
         2022-08-07&    59798.3450&     1.46&     1.35&  25& 40\\
\hline
\multicolumn{6}{c}{$K_s$-band}   \\
\hline
         2022-07-04&    59764.4255&     1.47&     1.32&  25& 46\\
         2022-07-04&    59764.4895&     1.60&     1.53&  25& 40\\
         2022-07-05&    59765.2637&     1.53&     1.46&  25& 40\\
         2022-07-05&    59765.3360&     1.61&     1.29&  25& 46\\
         2022-07-05&    59765.3971&     1.62&     1.28&  25& 40\\
         2022-07-05&    59765.4654&     1.61&     1.44&  25& 40\\
         2022-07-06&    59766.4436&     1.61&     1.38&  25& 40\\
         2022-08-07&    59798.3607&     1.38&     1.39&  25& 40\\
\enddata
\tablecomments{MJD: Modified Julian Date (JD$-$2,400,000.5). FWHM: Measured median full width at half maximum using {\texttt {SExtractor}}. AM: Median airmass. $N_\textrm{f}$: Number of dithered frames on target per epoch. ET: Exposure time (in seconds) for each dithered frame.}
\end{deluxetable}

The pre-processed science images were downloaded from the \texttt{IDL} Interpretor of the WIRCam Images (\texttt{$`$I$`$iwi}{\footnote{\url{https://www.cfht.hawaii.edu/Instruments/Imaging/WIRCam/IiwiVersion2Doc.html}}}) preprocessing pipeline at CFHT which incorporates detrending (dark subtraction, flat-fielding) and initial sky subtraction. We first used \texttt{WeightWatcher} \citep{marmo2008} to create weight maps for each pre-processed image thereby masking bad pixels in the WIRCam mosaic. These weight maps were used along with the images in \texttt{SExtractor} \citep{bertin1996} to generate a catalogue of sources matching those from the Two Micron All Sky Survey (2MASS) Point Source Catalog \citep{skrutskie2006}. These matched source catalogues were used to obtain astrometric solution for each pre-processed image using \texttt{SCAMP} \citep{bertin2006}. We note that the \texttt{SCAMP} also scales the flux of each detector with their magnitude zero-points and performs an internal photometric calibration against 2MASS. Finally with the astrometric solutions, the dithered images at a given epoch were median-combined using \texttt{SWARP} \citep{bertin2002} at the instrument pixel scale for photometric data reduction. These co-added median combined images for each filter independently and for each epoch were used as input for photometry of point sources.

\subsection{Point-spread function photometry}

The point-spread function (PSF) photometry on WIRCam images has been discussed in detail in \citet{bhardwaj2020a} and \citet{bhardwaj2021a} for Messier 3 and Messier 15 globular clusters. In brief, the standard \texttt{DAOPHOT/ALLSTAR} \citep{stetson1987} and \texttt{ALLFRAME} \citep{stetson1994} routines were used for PSF photometry. Initially, an aperture photometry was performed on all sources with brightness above $5\sigma$ of the detection threshold using \texttt{DAOPHOT}. An empirical PSF was generated from 200 bright, isolated, non-saturated stars excluding those around the corners of each detector. These PSFs were used to obtain \texttt{ALLSTAR} photometry for all sources in a given image. For the \texttt{ALLFRAME} run, a common reference-star list was obtained from a median-combined image using the best-seeing epoch in $JK_s$ filters (Aug 8, 2022, FWHM$\sim$1.4). The frame-to-frame transformations were derived between the reference star-list and all epoch images using \texttt{DAOMATCH/DAOMASTER}. These transformations together with the reference star-list were used for the PSF photometry in \texttt{ALLFRAME}, which performs profile fitting of all sources  across all the frames, simultaneously. The output photometry in all the frames were internally calibrated using 50 secondary standards present in each frame in a given NIR filter. These non-variable secondary standards were chosen to have small photometric variability and were used to correct for the epoch-dependent frame-to-frame zero-point variations obtaining final catalog with instrumental $JHK_s$ photometry. 

\begin{figure*}
\centering
  \begin{tabular}{@{}cc@{}}
\includegraphics[width=0.5\textwidth]{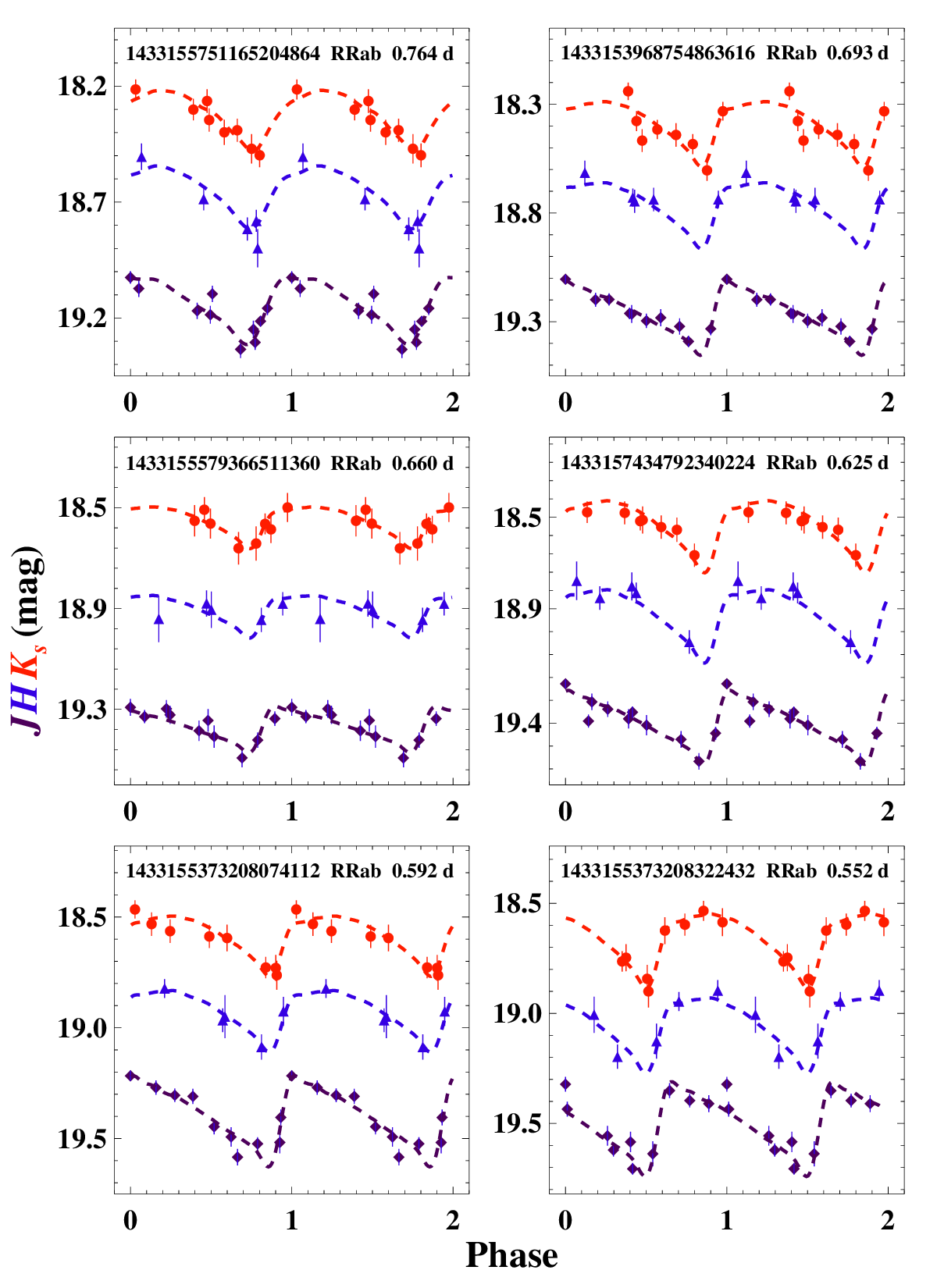} &
\includegraphics[width=0.5\textwidth]{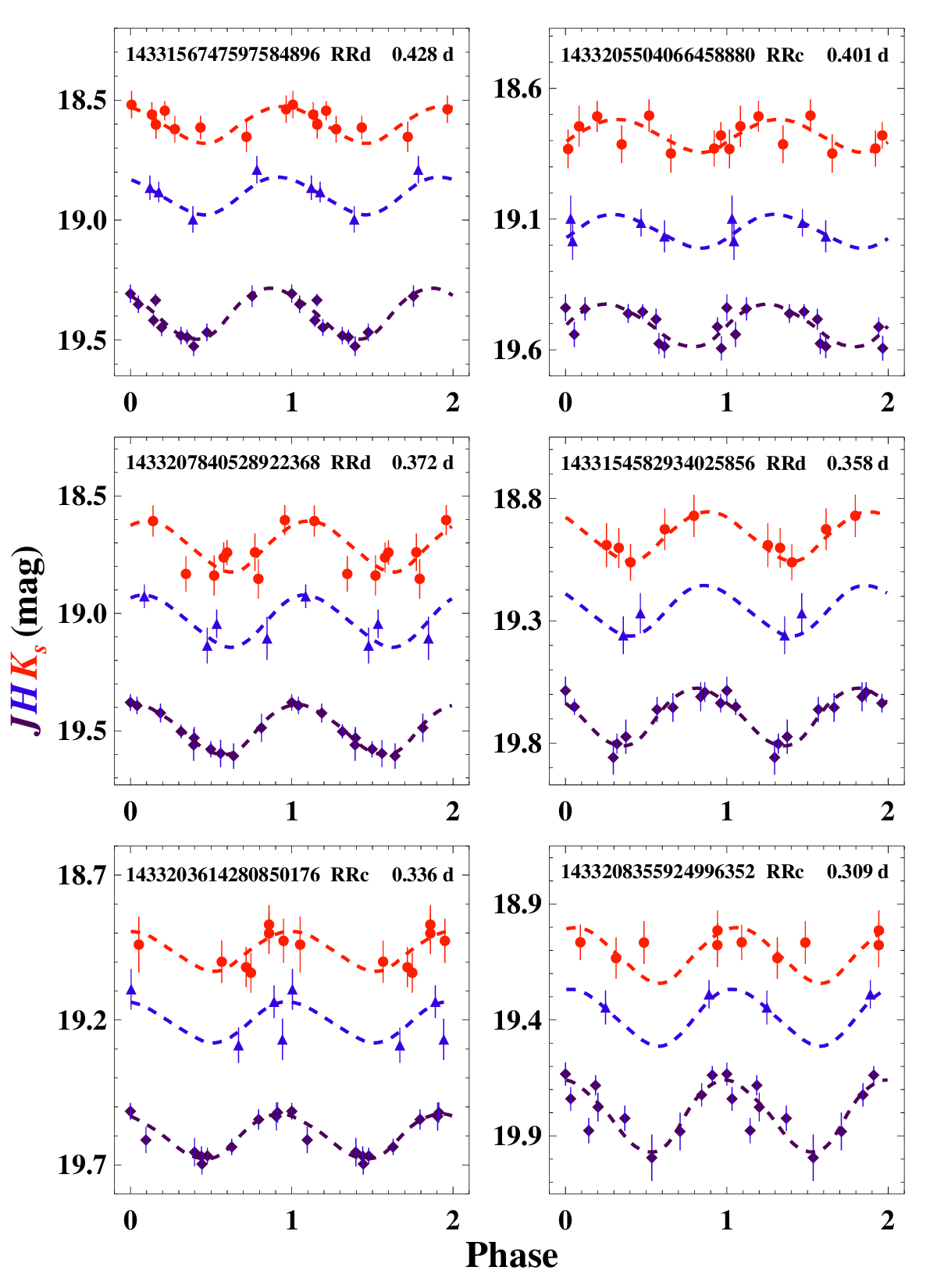} \\
  \end{tabular} 
  \caption{Example phased light curves of RR Lyrae stars in NIR bands covering the entire range of periods in our sample. {\it Left:} RRab stars. {\it Right:} RRc/RRd stars. The $J$-band (blue) and $K_s$-band (red) light curves are offset for clarity by $+0.2$ mag and $-0.3$~mag, respectively. The dashed lines represent the best-fitting templates to the data in each band. The {\it Gaia} DR3 star ID, variable subtype, and the pulsation period are included at the top of each panel.}
\label{fig:lcs_rrl}
\end{figure*}

The photometric calibration was performed by matching the instrumental photometry with 2MASS catalog within a search radius of $1.0\arcsec$. We found 556 common sources, which were restricted to 186 stars with the best-quality photometric flags in 2MASS, small uncertainties ($< 0.1$~mag), and spatial location away from the detector edges. The instrumental magnitudes were corrected for a fixed zero-point offset to convert into 2MASS system. The zero-point offsets in $JHK_s$ bands were determined accurately with uncertainties of 0.01 mag in all three filters. These 2MASS standards covered a typical color-range of $0.3 < (J-K_s) < 1.0$~mag, and no statistically significant variations were found in the zero-point offsets within this color range. A narrow color-range  ($\Delta(J-K_s) \lesssim$ 0.7~mag) and large uncertainties in colors (up to 0.15 mag) prevented an accurate and precise quantification of the color-coefficient in the transformation. Therefore, we did not apply color corrections. We estimate an uncertainty of 0.01 mag in photometry given the small color range of RRL and ACEP stars ($\Delta(J-K_s) \lesssim$ 0.5~mag).

\section{Variable stars in Draco}
\label{sec:var}

We used an initial list of 379 variable stars towards Draco dSph compiled by \citet{muraveva2020}. The compilation includes positions, periods, and variability types from several different literature studies and {\it Gaia} DR2 data \citep[see][for details]{muraveva2020}. This initial list included 336 RRL and 9 ACEP variables, which were cross-matched with our NIR photometry catalog within $1\arcsec$ tolerance. We recovered all 217 variables within the field of view of WIRCam consisting of 212 RRL -  173 fundamental mode (RRab), 23 first-overtone (RRc), and 15 mixed-mode (RRd) variables, and 5 ACEP variable stars. The light curves in $JHK_s$ bands were extracted for these 217 variables in Draco, which were phased using the adopted periods from the list of \citet{muraveva2020}. The mixed-mode (RRd) variables were phased with their dominant first-overtone periods. 

\subsection{Template-fitted light curves}

\begin{figure}
\centering
\includegraphics[width=0.48\textwidth]{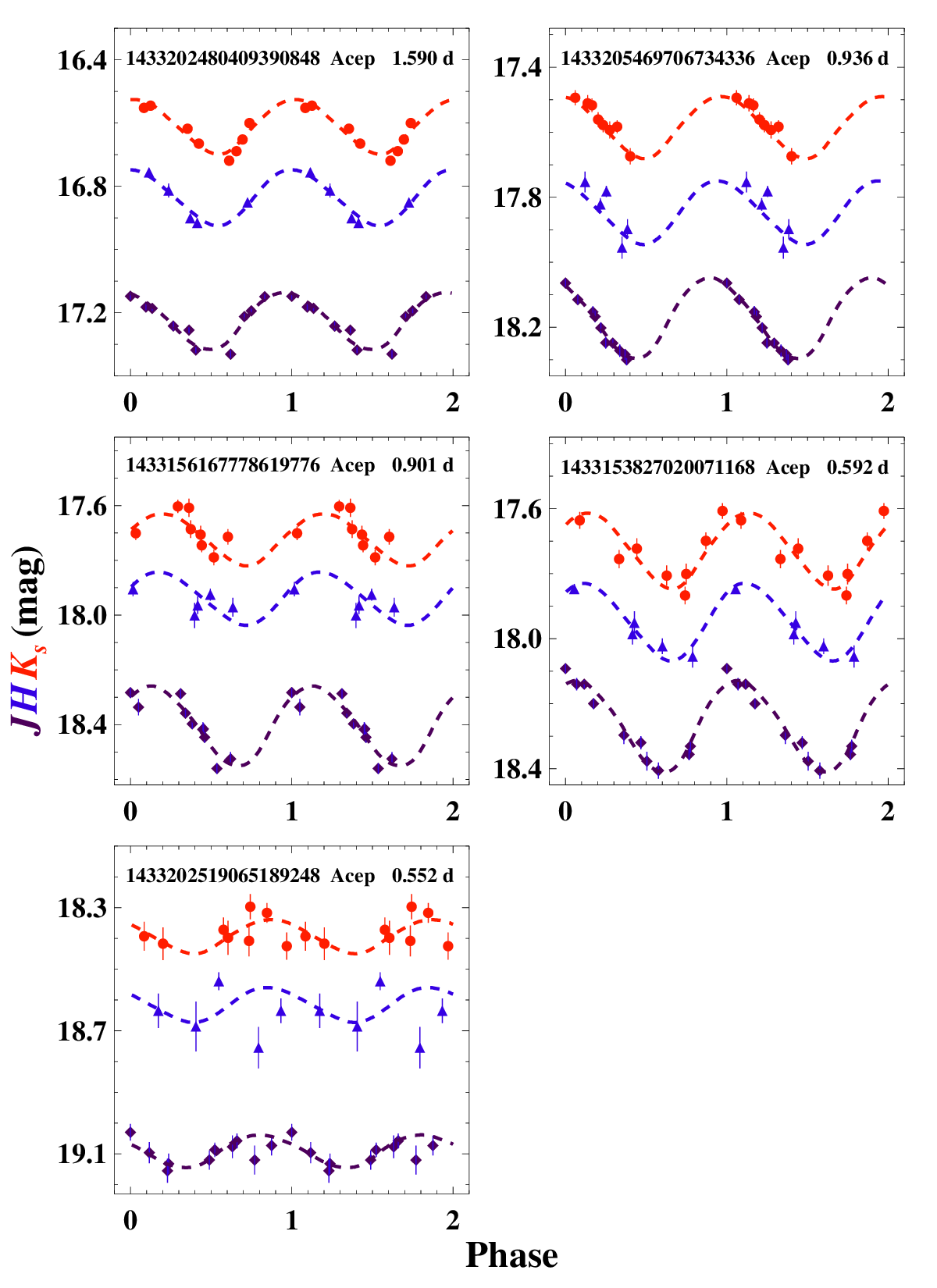}
	\caption{Same as Fig.~\ref{fig:lcs_rrl} but for ACEP stars. The $J$-band (blue) and $K_s$-band (red) light curves are offset for clarity by $+0.1$ mag and $-0.2$~mag, respectively.} 
\label{fig:lcs_acep}
\end{figure}

\begin{figure}
\centering
\includegraphics[width=0.96\columnwidth]{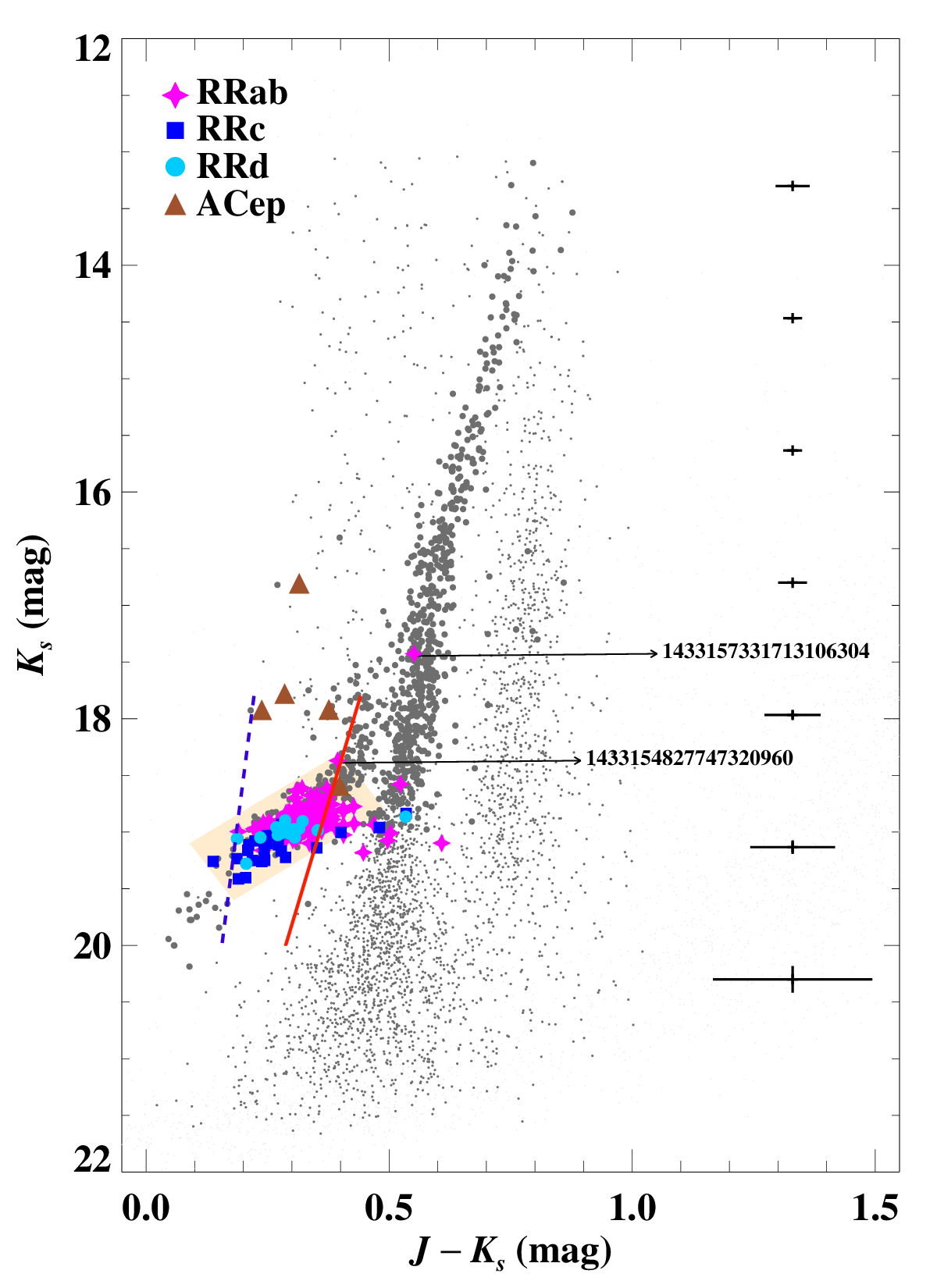}
	\caption{NIR color--magnitude diagram for Draco dSph in $(J-K_s),K_s$ (small grey dots). The larger symbols represent likely Draco members based on their proper-motions in {\it Gaia} DR3. Variable stars studied in this work are also overplotted with their mean magnitudes and colors. The theoretically predicted instability strip boundaries - fundamental red edge (solid red line) and first-overtone blue edge (dashed blue line) are taken from \citet{marconi2015}. On the right the representative $\pm3\sigma$ error bars are shown as a function of magnitude and color.} 
\label{fig:cmd_all}
\end{figure}

The phased light curves were fitted with NIR templates for RRL stars from \citet{braga2019}. We fitted all three sets of RRab templates from \citet{braga2019} to their observed NIR light curves irrespective of the periods. The $J$ band reference epoch corresponding to the maximum light was used to phase all three $JHK_s$ band light curves. Therefore, we solved for the mean-magnitude, peak-to-peak amplitude, and a phase offset in the case of $J$ band. The measured phase offset was kept fixed for the template-fitting in $HK_s$ bands. In the case of $K_s$ band,  mean-magnitudes and amplitudes were determined simultaneously. However, we only solved for the mean magnitudes in $H$ band assuming the amplitude to be similar to that in the $K_s$ band. This is a reasonable approximation since the light curve shape in $H$ and $K_s$ bands do not change significantly and there are not enough $H$ band epochs to independently constrain the amplitudes. 

Fig.~\ref{fig:lcs_rrl} displays template-fitted light curves of RR Lyrae stars in $JHK_s$ bands covering the entire period range. The phased light curves of RRab stars are well sampled in $JK_s$ bands while the templates reproduce the light curve variations of few epochs in the $H$ band. In the case of RRc/RRd, we use the single near-sinusoidal template provided by \citet{braga2019} for all stars. The sinusoidal set of $JHK_s$ templates was also used for the ACEP light curves since there are no NIR templates available for these stars. Fig.~\ref{fig:lcs_acep} shows template-fitted light curves for ACEP stars. The faint and low-amplitude RRc stars exhibit larger scatter in their light curves while the brighter ACEP show distinct and clear light curve variations. 
The best-fitting light curve templates for RRL and ACEP stars were used to derive their intensity-weighted mean magnitudes and colors. The median value of photometric uncertainties for the intensity-weighted mean magnitudes of RRL and ACEP are 0.036/0.045/0.034 mag and 0.030/0.034/0.022 mag in $J/H/K_s$, respectively. Table~\ref{tbl:var} lists the pulsation properties of variable stars in Draco dSph.

\subsection{Color--magnitude diagrams}
\label{subsec:cmd}

\begin{deluxetable*}{lccclcccccc}
\tablecaption{NIR pulsation properties of variable stars in Draco dSph. \label{tbl:var}}
\tablewidth{0pt}
\tablehead{
{ID} & {RA} & {Dec} & {P} &  {Type}& \multicolumn{3}{c}{Mean magnitudes ($m_\lambda$)}  & \multicolumn{3}{c}{$\sigma_{m_\lambda}$}  \\
 	&	&   &    &	   & $J$  &   $H$  & $K_s$  & $J$  &  $H$  & $K_s$  	\\  
 	&	deg.&	deg.	& days 	    &	   & \multicolumn{3}{c}{mag}  & \multicolumn{3}{c}{mag}}
  \startdata
                    1433161145644413696&   260.222875&    57.951278&    0.68171&      RRab&   19.126&   18.772&   18.760&    0.033&    0.046&    0.033\\
                    1433203889158772608&   259.898833&    57.975611&    0.68418&      RRab&   19.086&   18.799&   18.749&    0.037&    0.059&    0.028\\
                    1433153620861638528&   259.858833&    57.814250&    0.68939&      RRab&   19.113&   18.830&   18.794&    0.022&    0.069&    0.026\\
                    1433153968754863616&   259.840833&    57.855944&    0.69306&      RRab&   19.057&   18.758&   18.681&    0.026&    0.022&    0.035\\
                    1433204237051138304&   260.078958&    58.010528&    0.69485&      RRab&   19.094&   18.739&   18.780&    0.047&    0.061&    0.040\\
                    1433154789092499584&   260.087792&    57.871944&    0.73202&      RRab&   19.004&   18.663&   18.698&    0.020&    0.023&    0.028\\
                    1433203859095469440&   259.872750&    57.973806&    0.74363&      RRab&   19.015&   18.730&   18.676&    0.023&    0.028&    0.021\\
                    1433155751165204864&   260.025042&    57.896944&    0.76413&      RRab&   18.934&   18.644&   18.613&    0.033&    0.067&    0.039\\
                    1433161832839184512&   260.162458&    57.959000&    0.79061&      RRab&   18.988&   18.672&   18.616&    0.016&    0.040&    0.016\\
                    1433156167778619776&   260.074110&    57.952090&    0.90087&      Acep&   18.295&   17.936&   17.919&    0.031&    0.023&    0.022\\
                    1433205469706734336&   259.783690&    57.976370&    0.93649&      Acep&   18.064&   17.844&   17.779&    0.030&    0.044&    0.022\\
                    1433202480409390848&   259.900190&    57.904310&    1.59027&      Acep&   17.121&   16.833&   16.806&    0.023&    0.034&    0.027\\
\enddata
\tablecomments{The {\it Gaia} DR3 source ID are provided in the first column.\\
(This table is available in its entirety in machine-readable form.)}
\end{deluxetable*}

\begin{figure*}
\centering
\includegraphics[width=0.97\textwidth]{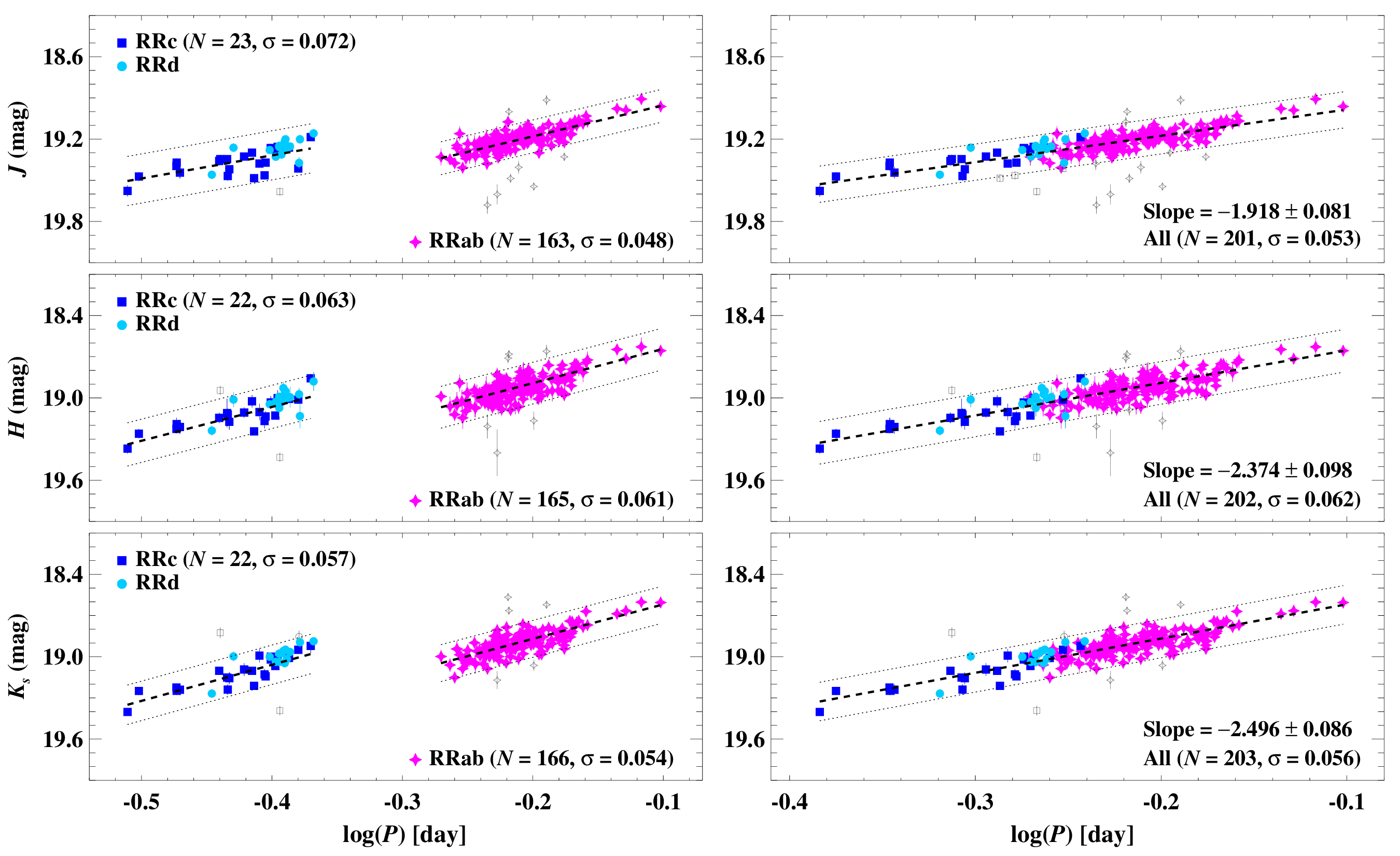}
\caption{Near-infrared period-luminosity relations for RRab and RRc stars (left) and all RRL stars (right) in $J$ (top), $H$ (middle), and $K_s$ (bottom). In the right panels, the periods for the RRc stars have been shifted to their corresponding fundamental-mode periods, as explained in the text. The grey symbols represent outliers removed iteratively during the fitting procedure. The dashed lines represent best-fitting linear regressions over the period range under consideration, while the dotted lines display $\pm 2.5\sigma$ offsets from the best-fitting PLRs.} 
\label{fig:rrl_plr}
\end{figure*}

Fig.~\ref{fig:cmd_all} displays NIR color--magnitude diagram for Draco dSph galaxy based on $\sim 4000$ point-like sources. The magnitudes and colors for all sources were corrected for extinction adopting a color-excess value of $E(B-V)=0.027$ mag \citep{schlegel1998} following previous works by \citet{muraveva2020} and \citet{kinemuchi2008}. We adopted the reddening law of \citet{card1989} and determined extinction in NIR using $A_{J/H/K_s} = 0.94/0.58/0.39 E(B-V)$. \citet{bhardwaj2023} adopted the same reddening coefficients to provide calibrated NIR period-luminosity-metallicity (PLZ) and period-Wesenheit-metallicity (PWZ) relations for RRL stars which will be used for the distance measurement in Section~\ref{sec:dis}.

The sources in the color--magnitude diagram were also cross-matched with the {\it Gaia} DR3 catalogue \citep{vallenari2022} to extract proper motions of $\sim 2300$ stars. The majority of fainter stars ($K_s > 19$ mag) do not have proper motions. We found mean proper motions along the right ascension and declination axes to be $0.027\pm0.014$ and $-0.220\pm0.014$ mas/yr with a scatter of 1.02 and 0.94 mas/yr, respectively after restricting the sample to good astrometry (renormalized unit weight error $< 1.4$) and photometry ($\sigma_{JHK_s} < 0.3$ mag). These mean proper motions are in good agreement with previous determinations based on {\it Gaia} data \citep[e.g., $0.044\pm0.005$ and $-0.188\pm0.006$ mas/yr in][]{pace2022}. The proper-motion cleaned color--magnitude diagram (in larger grey symbols) clearly exhibits well-populated horizontal and red giant branches. 

The distribution of variable stars in the color--magnitude diagram is also shown in Fig.~\ref{fig:cmd_all}. The abundant RRL population is distinctly evident at the horizontal branch. The predicted boundaries of the instabillity strip were taken from \citet{marconi2015} based on RRL pulsation models. The instability strip boundaries were offset by a distance modulus of $19.53\pm0.07$ mag \citep{muraveva2020}. Majority of RRL stars occupy the location between the predicted fundamental red edge and the first-overtone blue edge. One of the brightest RRL (DR3-1433157331713106304) does not show any variability in its light curve within the photometric errors while NIR light curves of DR3-1433154827747320960 exhibit large scatter. 
Most variables located in the color--magnitude diagram beyond the predicted edges of the instability strip are still consistent with expectations. These variables exhibit larger scatter in $J$ and/or $K_s$ light curves, and their colors are within $1\sigma$ scatter from the instability strip boundaries. \\

% 1433155957325735040 SXP
% 1433203889158778880 Var (very red j-k=1.66)
% 1433155991683428864 Var (redder, clear periodicity in J)
% 1433157331713106304 RRab low amplitude

\begin{figure}
\centering
\includegraphics[width=0.97\columnwidth]{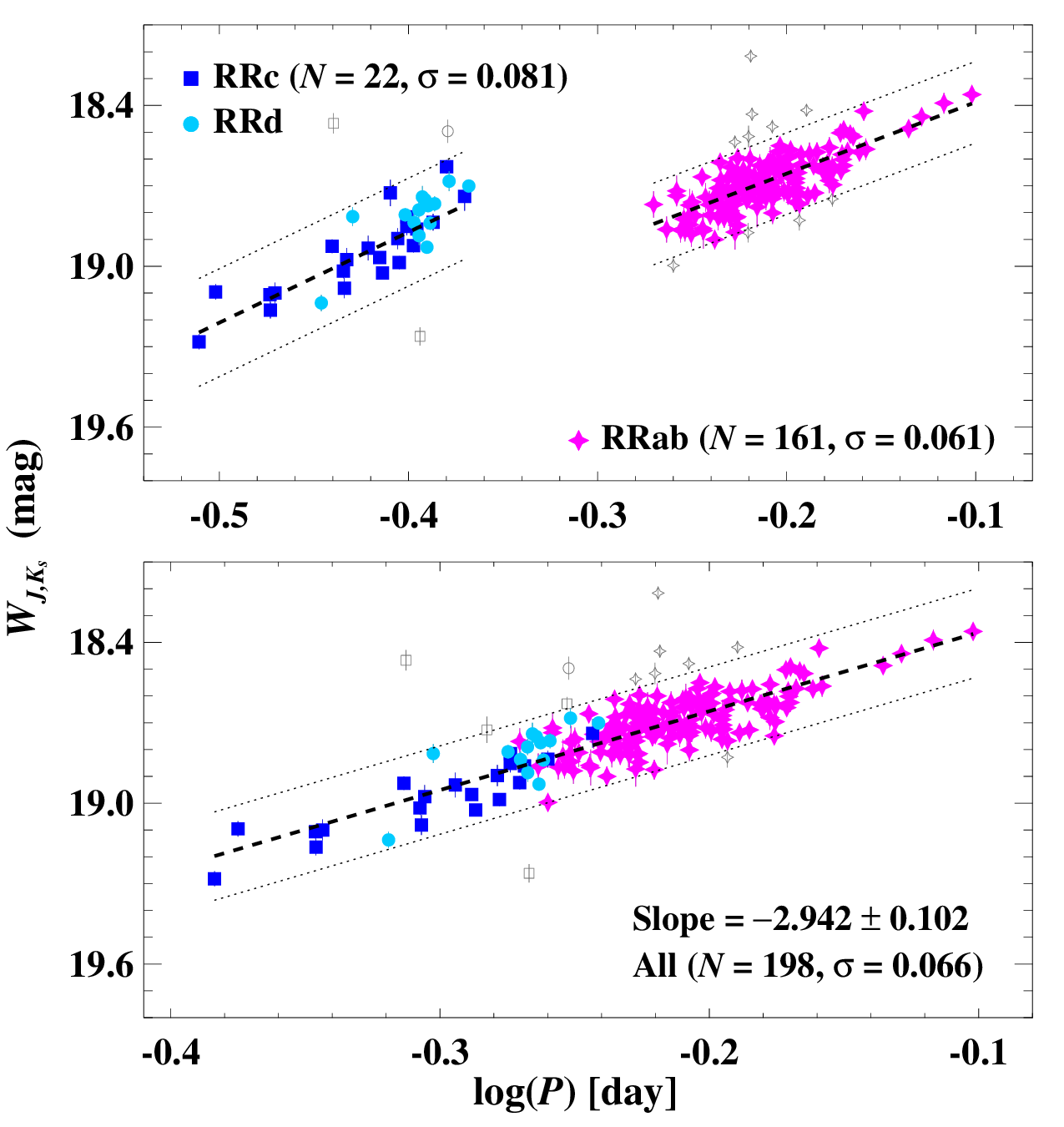}
\caption{Same as Fig.~\ref{fig:rrl_plr}, but for $W_{J,K_s}$ period-Wesenheit relation for RRab and RRc stars (top) and all RRL stars (bottom).} 
\label{fig:rrl_pwr}
\end{figure}

\section{Variable stars as distance indicators}
\label{sec:plrs}

\subsection{RR Lyrae PL and PW relations}

The extinction-corrected intensity-weighted mean magnitudes for RRL stars were used to derive PL relations in NIR bands for the first time in Draco dSph. Three different samples of RRL stars were considered: 
(1) RRab variables only, (2) RRc stars only, and (3) all RRL stars where the dominant first-overtone periods were used for RRd stars. The first-overtone periods were fundamentalized using $\log(P_\textrm{RRab})=\log(P_\textrm{RRc/RRd})+0.127$ \citep{coppola2015, braga2022}. Assuming that the PL relations are linear over the entire period range, the following relation was fitted to the data:

\begin{eqnarray}
m_\lambda &=& a_\lambda + b_\lambda \log(P),
\label{eq:plr}
\end{eqnarray}

\noindent where $m_\lambda$ is the extinction-corrected mean magnitude in $JHK_s$ bands, and $a_\lambda$ and $b_\lambda$ give the slope and zero-point of the PL relation. We fitted a linear regression iteratively removing the single largest $2.5\sigma$ residual of the PL relation under consideration.

Fig.~\ref{fig:rrl_plr} displays PL relations for RRL stars in $JHK_s$ bands. It is evident that the scatter in the PL relations for all RRL sample is small ($\sim0.05$ mag) and comparable to those seen in globular clusters \citep{bhardwaj2023}. The scatter typically results from the intrinsic width of the instability strip, a possible metallicity scatter, and photometric uncertainties on mean-magnitudes. It can be noted that the dispersion in the $H$-band PL relation is relatively larger than in $JK_s$ bands because the mean-magnitudes in $H$ band are based on a smaller number of epochs. Since the metallicity coefficients of RRL PLZ relations in the NIR are of the order of $\sim 0.2$ mag/dex \citep{marconi2015, bhardwaj2023}, the small dispersion of $\sim0.05$ mag in NIR PL relations suggests no statistically significant scatter in [Fe/H] for RRL in Draco dSph. If we assume the intrinsic width of the instability strip of $0.03$~mag \citep[e.g., RRL PL relations in M53,][]{bhardwaj2021} and attribute the remaining scatter of $0.04$~mag to metallicity, then the dispersion in [Fe/H] is expected to be $\sim0.2$~dex in these centrally located RRL stars.

Table~\ref{tbl:plr_rr} lists PL relations for different samples of RRL stars. The slopes of PL relations become steeper moving 
from $J$ to $K_s$, as typically seen for RRL stars. The slopes are best-constrained for the global sample of all RRL stars. 
The slopes of RRL PL relations in Draco dSph are in good agreement with the slopes of NIR PLZ relations for RRL in globular clusters \citep[$-1.83\pm0.02$ in $J$, $-2.29\pm0.02$ in $H$, and $-2.37\pm0.02$ in $K_s$,][]{bhardwaj2023} within their $1.5\sigma$ uncertainties. 

We also derived PW relations for RRL stars in Draco dSph. The following reddening-free Wesenheit magnitudes \citep{madore1982} were derived in NIR filters:  $W_{J,H} = H - 1.63(J-H)$, $W_{J,K_s} = K_s - 0.69(J-Ks)$, $W_{H,K_s} = K_s - 1.92(H-K_s)$, similar to \citet{bhardwaj2023a}. These Wesenheit magnitudes were used in eqn.~(\ref{eq:plr}) to derive PW relations, and the results of the best-fitting linear regression are listed in Table~\ref{tbl:plr_rr}. Fig.~\ref{fig:rrl_pwr} displays $W_{J,K_s}$ PW relation for RRL stars, which is best-constrained among these empirical relations. The scatter in the PW relations is larger than PL relations, in particular, in the Wesenheit magnitudes involving $H$-band. This is due to larger uncertainties in $J-H$ and $H-K_s$ colors because of lower number of epochs used to determine $H$-band RRL mean magnitudes. The larger color-coefficients and the uncertainties in colors contribute to the significantly increased scatter in $W_{J,H}$ and $W_{H,K_s}$ PW relations as comapred to PL relations. Nevertheless, the slopes of PW relations are in agreement with those of RRL in globular clusters \citep{bhardwaj2023a}.

\begin{deluxetable}{cccccc}
\tablecaption{Near-infrared PL and PW relations of RRL stars in Draco dSph. \label{tbl:plr_rr}}
\tablewidth{0pt}
\tablehead{
{Band} & {Type} & {$a_\lambda$} & {$b_\lambda$} & {$\sigma$}& {$N$}}
\startdata
     $J$ &  RRab &    18.728$\pm$0.024      &$     -2.260\pm0.118      $&      0.048 &  163\\
         &   RRc &    18.645$\pm$0.183      &$     -1.684\pm0.437      $&      0.072 &   23\\
         &   All &    18.793$\pm$0.018      & $    -1.918\pm0.081      $&      0.053 &  201\\
     $H$ &  RRab &    18.393$\pm$0.033      &$     -2.499\pm0.161      $&      0.061 &  165\\
         &   RRc &    18.079$\pm$0.158      &$     -2.465\pm0.374      $&      0.063 &   22\\
         &   All &    18.416$\pm$0.022      &$     -2.374\pm0.098      $&      0.062 &  202\\
   $K_s$ &  RRab &    18.368$\pm$0.028      &$     -2.511\pm0.135      $&      0.054 &  166\\
         &   RRc &    18.007$\pm$0.141      &$     -2.629\pm0.330      $&      0.057 &   22\\
         &   All &    18.368$\pm$0.019      &$     -2.496\pm0.086      $&      0.056 &  203\\
   \hline
    $W_{J,H}$ &  RRab &    17.822$\pm$0.046      &$     -3.074\pm0.226      $&      0.092 &  156\\
              &   RRc &    17.567$\pm$0.290      &$     -2.742\pm0.684      $&      0.094 &   20\\
              &   All &    17.858$\pm$0.031      &$     -2.880\pm0.139      $&      0.094 &  191\\
  $W_{H,K_s}$ &  RRab &    18.327$\pm$0.048      &$     -2.440\pm0.232     $ &      0.085 &  156\\
              &   RRc &    17.754$\pm$0.217      &$     -3.169\pm0.515     $ &      0.097 &   20\\
              &   All &    18.254$\pm$0.032      &$    -2.814\pm0.145      $&      0.090 &  192\\
  $W_{J,K_s}$ &  RRab &    18.121$\pm$0.032      &$     -2.667\pm0.153      $&      0.061 &  161\\
              &   RRc &    17.522$\pm$0.193      &$     -3.377\pm0.452      $&      0.081 &   22\\
              &   All &    18.069$\pm$0.023      &$     -2.942\pm0.102      $&      0.066 &  198\\    
\enddata
\tablecomments{The zero-point ($a$), slope ($b$), dispersion ($\sigma$) and the number of stars ($N$) in the final PL fits are tabulated.}
\end{deluxetable}

\subsection{Anomalous Cepheid PL and PW relations}

We also investigated NIR PL relations for a small sample of 5 ACEP variables in Draco dSph. Previously, $K_s$-band PL relations for ACEP have been investigated only in the Large Magellanic Cloud \citep[LMC,][]{ripepi2014a}, and no other empirical PL relations are available for these variables at NIR wavelengths. \citet{ripepi2014a} derived $K_s$-band PL relations for both fundamental and first-overtone mode ACEP variables. Therefore, to identify the pulsation mode of ACEP in Draco dSph, we compared the location of their PL relations with ACEPs in the LMC. The distance moduli $\mu_\textrm{Draco}=19.53\pm0.07$~mag \citep{muraveva2020} and $\mu_\textrm{LMC}=18.48\pm0.03$ ~mag \citep{piet2019} were adopted to calibrate the zero-points of corresponding ACEP PL relations for a relative comparison.

\begin{figure}
\centering
\includegraphics[width=0.97\columnwidth]{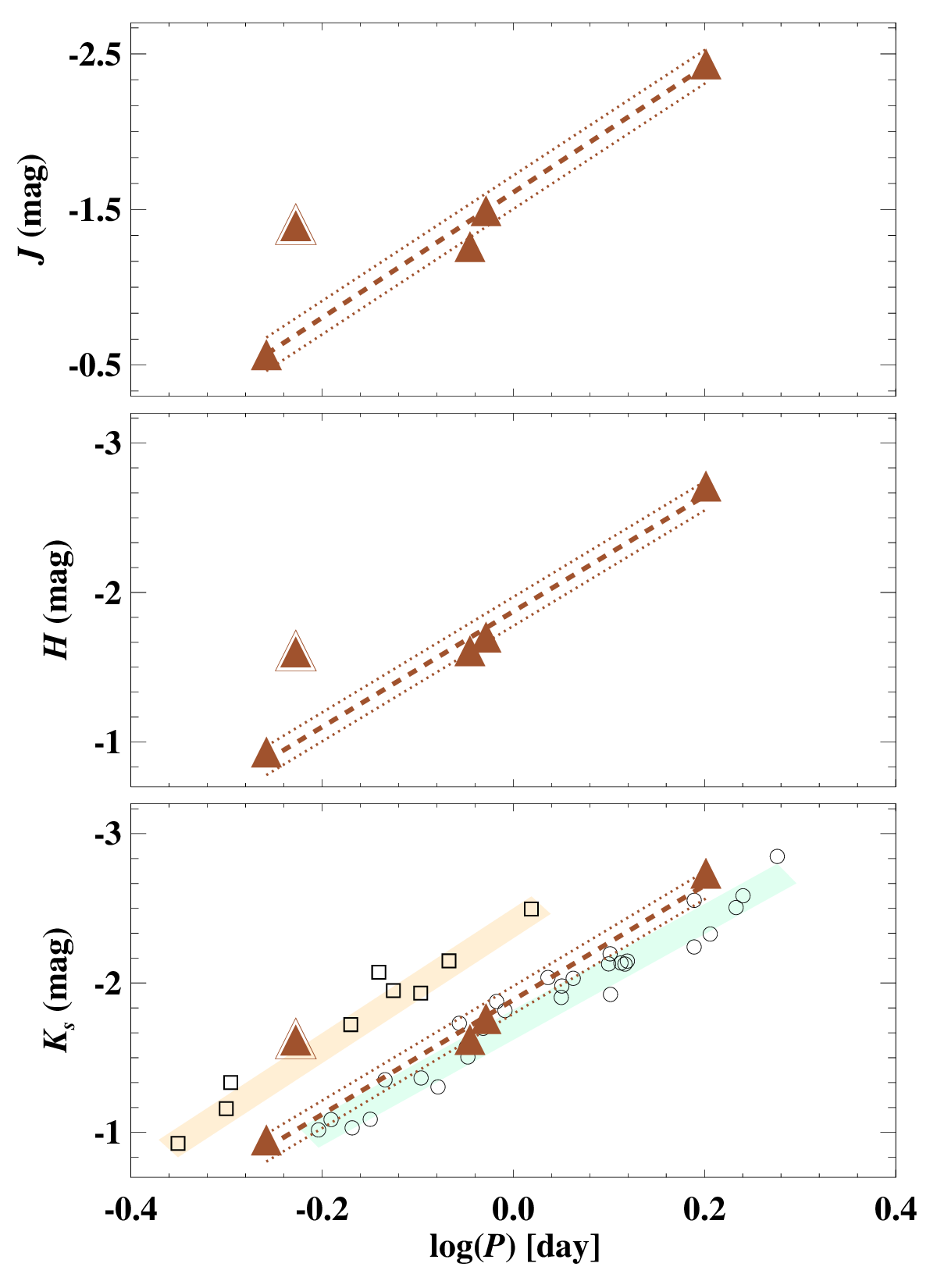}
\caption{Near-infrared period-luminosity relations for ACEP (triangles) in $J$ (top), $H$ (middle), and $K_s$ (bottom). In the bottom panel, open circles and squares represent fundamental and first-overtone mode ACEP in the LMC, respectively, and the shaded regions represent their PL relations with $\pm1\sigma$ scatter \citep{ripepi2014a}. The dashed-lines represent best-fitting PL relation to four ACEP pulsating in the fundamental mode while the dotted lines display $\pm 2.5\sigma$ offsets from the best-fitting relation. The first-overtone mode ACEP is marked with overplotted open triangle.} 
\label{fig:acep_plr}
\end{figure}

The bottom panel of Fig.~\ref{fig:acep_plr} displays $K_s$-band PL relation for ACEP in Draco dSph and the LMC. The figure shows that four ACEP in our sample occupy the location along the PL relation for fundamental mode variables while one ACEP is an overtone variable. The first-overtone ACEP (DR3-1433153827020071168) was also found to follow $V$-band PL relation of overtone mode pulsators in \citet{kinemuchi2008}, but the larger scatter in optical bands precluded authors to confirm its sub-classification. The tight PL relations in NIR allow us to confirm that this ACEP is pulsating in the first-overtone mode. The four fundamental mode ACEP fall along a linear sequence and cover a period range of $\Delta \log(P)\sim0.45$~days. This period range is sufficient to derive PL relations in $JHK_s$ bands which are also shown in Fig.~\ref{fig:acep_plr}. The results of the best-fitting PL relations in the form of equation~(\ref{eq:plr}) are tabulated in Table~\ref{tbl:plr_ac}. Despite the small sample size, the slope of ACEP PL relation in the $K_s$-band agrees well with the slope of fundamental-mode PL relation in the LMC \citep[$-3.54\pm0.15$,][]{ripepi2014a}.

\begin{figure}
\centering
\includegraphics[width=0.97\columnwidth]{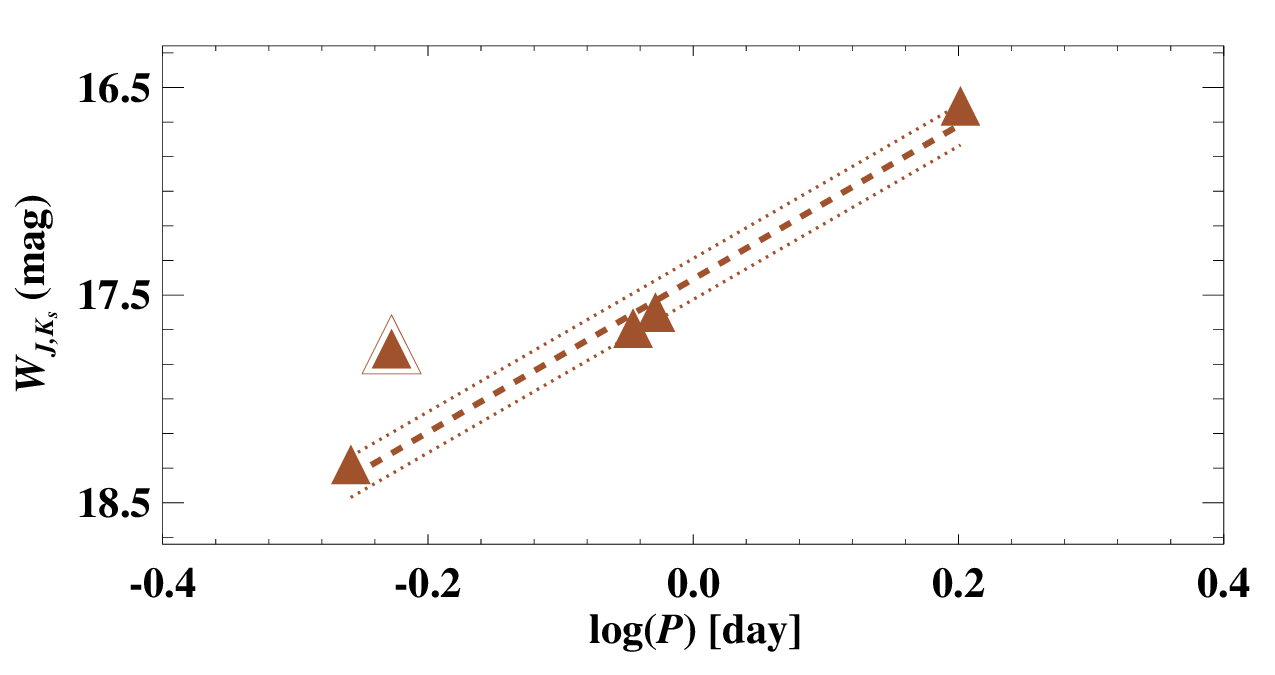}
\caption{Same as Fig.~\ref{fig:acep_plr}, but for $W_{J,K_s}$ period-Wesenheit relation for ACEP stars.} 
\label{fig:acep_pwr}
\end{figure}

\begin{deluxetable}{cccccc}
\tablecaption{Near-infrared PL and PW relations of ACEP stars in Draco dSph. \label{tbl:plr_ac}}
\tablewidth{0pt}
\tablehead{
{Band} & {Type} & {$a_\lambda$} & {$b_\lambda$} & {$\sigma$}& {$N$}}
\startdata
$  J $& ACEP-F&   17.940$\pm$0.031   &$   -4.029\pm0.149   $&   0.108&  4\\
$  H $& ACEP-F&   17.677$\pm$0.037   &$   -3.865\pm0.231   $&   0.097&  4\\
$ K_s$& ACEP-F&   17.643$\pm$0.034   &$   -3.832\pm0.207   $&   0.092&  4\\
\hline
$  W_{J,H} $& ACEP-F&   17.145$\pm$0.057   &$   -3.464\pm0.272   $&   0.186&  4\\
$  W_{H,K_s} $& ACEP-F&   17.615$\pm$0.044   &$   -3.901\pm0.275   $&   0.112&  4\\
$ W_{J,K_s}$& ACEP-F&   17.422$\pm$0.036   &$   -3.692\pm0.218   $&   0.099&  4\\
\enddata
\tablecomments{The zero-point ($a$), slope ($b$), dispersion ($\sigma$) and the number of stars ($N$) in the final PLR fits are tabulated.}
\end{deluxetable}

The calibrated zero-points of fundamental-mode ACEP PL relations in the Draco dSph and the LMC are systatically different as seen in the bottom panel of Fig.~\ref{fig:acep_plr}. With above-mentioned distance to the Draco dSph, the absolute zero-point of $-1.887\pm0.078$~mag is 0.17 magnitude brighter than that of ACEP in the LMC \citep{ripepi2014a}. We note that the slopes of these relations also differ by $\sim0.3$~mag/dex, but are consistent within $1.2\sigma$ of their combined uncertainties. Recently, the first spectroscopic observations of Galactic ACEP stars by \citet{ripepi2024} have confirmed that these are metal-poor stars ([Fe/H]$<-1.5$~dex), but their metallicities are not known in the LMC and Draco dSph. For RRL stars, the mean [Fe/H] values in the LMC \citep[$\sim-1.5$~dex,][]{gratton2004, borissova2009} and the Draco dSph \citep[$\sim-2.0$~dex,][]{muraveva2020} are comparatively different with the latter being more metal-poor. Similar to RRL stars, the mean metallicities of ACEP in the LMC and Draco dSph are also expected to be significantly different. Therefore, the difference in the zero-points may hint at a possible metallicity effect on their PL relations such that the more metal-poor stars are fainter. Assuming a metallicity difference of $\Delta$[Fe/H]$=0.5$~dex between LMC and Draco, a zero-point offset of 0.17 mag would imply a negative metallicity term of the order of $\sim-0.34$ mag/dex. This value falls in the range of metallicity coefficients of NIR PL relations for more popular classical Cepheids in several recent studies \citep[e.g.,][]{bhardwaj2023a, trentin2023a}.  

\begin{deluxetable*}{lllll}
\tablecaption{Distance modulus estimates to Draco dSph galaxy. \label{tbl:mu}}
\tablewidth{0pt}
\tablehead{
{$\mu$ (mag)} & {[Fe/H] (dex)} & {$E(B-V)$~(mag)} & {Method} & {Reference}}
\startdata
19.55$\pm$0.03  &  $-1.98\pm$0.10	&	0.027    &	RRL PLZ$_{K_s}$ relation	&TW	\\
19.55$\pm$0.03  &  $-1.98\pm$0.10	&	0.027    &	RRL PLZ$_H$ relation		&TW	\\
19.64$\pm$0.04  &  $-1.98\pm$0.10	&	0.027    &	RRL PLZ$_J$ relation 		&TW	\\
19.43$\pm$0.06  &   ~--- 	          &	0.027    &	ACEP PL$_{K_s}$ relation	&TW	\\
19.39$\pm$0.03$^a$	&	$-2.00\pm$~---  &	~---    &   PWZ$_{VI}$ relation			   &\citet{nagarajan2022}	\\
19.53$\pm$0.07	&	$-1.98\pm$0.01	&	0.027     & RRL  $M_G$--[Fe/H] relation		& \citet{muraveva2020}	\\
19.35$\pm$0.12	&	~---		    &	 ~---       &   Density model fitting			&\citet{hernitschek2019}	\\
19.51$\pm$0.03$^a$	&	~---		    &	~---      &   PLZ$_i$ relation			      &\citet{sesar2017}	\\
19.58$\pm$0.15	&	$-2.19\pm$0.03	&	0.027     & RRL  $M_V$--[Fe/H] relation		& \citet{kinemuchi2008}	\\
19.49$\pm$--	&	~---		    &	0.027     & ACEP PL$_V$ relation		    & \citet{kinemuchi2008} \\
19.49$\pm$0.15	&	~---		    &	0.03     & TRGB		                          &\citet{cioni2005}	\\
19.40$\pm$0.15	&	$-2.00\pm$--    &	0.027     & RRL  $M_V$--[Fe/H] relation		&\citet{bonanos2004}	\\
19.84$\pm$0.14	&	~---		    &	0.03     &	$M_V$ Horizontal Branch stars 	& \citet{bellazzini2002}	\\	%E(B − V) = 0.02 R = 3.1&
19.5$\pm$0.2	&	~---		    &	0.03     &	$M_V$ Horizontal Branch stars 	& \citet{aparicio2001}	\\	%E(B − V) = 0.02 R= 3.1&
19.6$\pm$0.3	&	~---		    &	0.03     &	$M_V$ RRd stars 				& \citet{nemec1985b}	\\	%E(B − V) = 0.02 R = 3.1&
19.4$\pm$0.3	&	~---		    &	0.03     &	$M_V$ Horizontal Branch stars 	& \citet{stetson1979}	\\	%E(B − V) = 0.02 R = 3.1&
\hline
\multicolumn{4}{l}{$\mu_\textrm{Draco} = 19.557~\pm~0.026$~(stat.)~$\pm~0.031$~(syst.)~mag}	&TW\\
\hline
\multicolumn{4}{l}{$D_\textrm{Draco} = 81.545~\pm~0.977$~(stat.)~$\pm~1.165$~(syst.)~kpc}	&TW\\
\enddata
\tablecomments{We adopted an uncertainty of 0.1 dex in mean metallicity from \citet{kirby2013}.\\
$^a$See the text in Section~\ref{sec:dis}.}
\end{deluxetable*}

Fig.~\ref{fig:acep_pwr} displays the $W_{J,K_s}$ band PW relation for ACEP variables. The coefficients of all Wesenheit relations are tabulated in Table~\ref{tbl:plr_ac}. Similar to RRL stars, the scatter increases in the PW relations in $W_{J,H}$ and $W_{H,K_s}$ bands. We note that there are no NIR PW relations for ACEP stars in the literature for a relative comparison.

\section{Distance to Draco dSph using population II stars}
\label{sec:dis}

Draco is one of the nearest satellites of the Milky Way. Yet, its distance is not very accurate despite several detailed studies of resolved stellar populations. Table~\ref{tbl:mu} lists distance moduli measurements to Draco dSph available in the literature. It is evident that the distance modulus varies between $19.35\pm0.03$~mag and $19.84\pm0.14$~mag in studies based on modern photometric data. The majority of these measurements are based on visual magnitude of horizontal branch stars or traditional $M_V-$[Fe/H] relation for RRL stars. These empirical relations suffer from evolutionary effects, and extinction and metallicity uncertainties, as mentioned in the Section~\ref{sec:intro}, and therefore exhibit larger realistic uncertainties of $0.1-0.2$ mag. Regardless, the resulting distance moduli typically range between 19.50 and 19.60 mag, except for the measurement from \citet{bellazzini2002}, which is at the high end with 19.84 mag. Two recent distance moduli estimates based on RRL density model fitting \citep{hernitschek2019} and optical PWZ relation \citep{nagarajan2022} are smaller than 19.40 mag. \citet{nagarajan2022} employed global fits to heterogeneous optical data for RRL stars and quoted a rather small statistical uncertainty of 0.03~mag, which presumably does not include any calibration errors since the uncertainty on their adopted PWZ zero-point itself is $\sim 0.05$~mag. Similarly, \citet{sesar2017} quoted a small uncertainty of only $0.03$~mag on distance modulus. The authors assumed a Gaussian metallicity distribution centered on $-1.5$~dex with a standard deviation of 0.3 dex for the RRL stars, which is different from most other studies, including this work. The quoted error only includes adopted error on absolute $i$-band magnitudes, excluding possible reddening and metallicity uncertainties in the calibration.

\subsection{RRL and ACEP variables}

We utilized accurate and precise PL relations for RR Lyrae stars in Draco dSph to determine its distance modulus. The calibrator PLZ relations in NIR bands were adopted from \citet{bhardwaj2023}, who provided the most precise empirical determination of the metallicity coefficient using homogeneous NIR photometry of $\sim 1000$ RRL in 11 globular clusters. The authors found that these empirical PLZ relations are in good agreement with theoretical predictions \citep{catelan2004, marconi2015}. The adopted calibrations and the observations presented in this work were obtained with the same instrument/telescope and with similar methodology, thus mitigating any possible analysis systematic uncertainties. 

To apply these empirical and theoretical calibrations, we need to assume a mean metallicity for RRL stars in Draco. \citet{muraveva2020} identified 16 RRL variables with spectroscopic metallicity measurements from \citet{walker2015} with a mean metallicity of $-1.98\pm0.65$~dex. \citet{walker2015} derived [Fe/H] values of red giant and horizontal branch stars in Draco with a median uncertainty of 0.20 dex, but the errors range between 0.06 and 0.98 dex for bright and faint stars. The large scatter of 0.65 dex in the mean metallicity of 16 RRL stars is likely due to larger uncertainties in the individual [Fe/H] measurements of horizontal branch stars. Nevertheless, this mean [Fe/H] value is in agreement with the mean metallicity of Draco of $-1.98\pm0.01$~dex ($\sigma=0.42$~dex) from \citet{kirby2013} based on spectroscopic observations of its members. Recently, \citet{taibi2022} derived a metallicity of $-1.86$ for Draco mainly using the red giant branch stars. Since the dispersion in NIR PL relations suggests a small scatter in metallicity, we adopted a mean metallicity of $-1.98$ \citep{kirby2013} with assumed error of $0.10$~dex for Draco dSph.

The distance modulus measurements based on RRL PL relations are listed in Table~\ref{tbl:mu}. In the $HK_s$-band, the distance moduli are $19.55\pm0.03$~mag with a systematic error of $0.03$~mag. The distance modulus is $\sim2\sigma $ larger in the $J$-band. We note that the absolute zero-points in the $J$-band differ by $0.06$~mag between empirical and theoretical calibrations \citep{bhardwaj2023a}, while they agree within $0.02$~mag in the $HK_s$ bands. The larger uncertainties in the $J$-band PLZ calibrations may be contributing to a relatively larger distance modulus. The uncertainties on the coefficients of the calibrator PLZ relations and the empirical PL relations were propagated to determine statistical errors. The systematic errors include the uncertainties on the zero-points of the calibrator PLZ relation ($\sim0.02-0.03$~mag) and empirical PL relation ($\sim0.02$~mag), and the uncertainty due to $0.1$~dex variation in the mean metallicity or metallicity scale ($\sim0.02$~mag). The errors due to reddening are negligible ($<0.005$~mag) considering small color-excess value and extinction coefficients in the NIR. 

We also determined RRL based distances adopting independent theoretical PLZ calibrations from \citet{marconi2015} in NIR bands. These distance moduli derivations agree well within $1\sigma$ with the quoted values in Table~\ref{tbl:mu} based on empirical calibrations. Due to larger scatter in the Wesenheit relations for Draco RRL and their empirical calibrations, we do not use those to determine distance modulus. Excluding relatively larger $J$-band determinations, we adopt a weighted average of distance moduli based on $HK_s$ band for both empirical and theoretical calibrations. Our recommended distance modulus to Draco dSph is $19.557~\pm~0.026$~(stat.)~$\pm~0.031$~(syst.) mag, which translates to a distance of $81.55~\pm~0.98$~(stat.)~$\pm~1.17$~(syst.) kpc. 

Our small sample of 5 ACEPs can also be used to determine an independent distance to Draco dSph. Only fundamental mode ACEPs were used for this analysis excluding one of the overtone mode variable. The empirically calibrated $K_s$-band PL relation for ACEPs in the LMC \citep{ripepi2014a} was used to determine absolute magnitude for Draco variables. 
The resulting distance modulus of $19.43\pm0.06$~mag based on ACEPs is also listed in Table~\ref{tbl:mu}. This distance modulus is smaller than the adopted RRL based determination, but is in agreement within $2\sigma$ of their quoted uncertainties. 
There are no other absolute calibrations of PL and PW relations in the NIR bands for ACEP stars. 

Thanks to unprecedented NIR photometry, our RRL distance measurements are significantly more precise and accurate, but also agree well with most previous determinations in the literature. The adopted calibrations and the observations presented in this work were obtained with the same telescope, same instrument and filters, as well as the same methodology applied for data reduction and measurements analysis, thus minimising any possible systematic uncertainties.

\subsection{Tip of the red giant branch}

Among population II distance indicators, the tip of the red giant (TRGB) is significantly brighter than RRL stars, and is an excellent distance indicator. In particular, the $I$-band magnitude of TRGB is remarkably constant over a range of metallicities, but their magnitude distribution is slanted or sloped with increasing metallicity at other wavelengths \citep{lee1993, rizzi2007, serenelli2017, freedman2020}. \citet{cioni2005} measured a TRGB-based distance of $19.49\pm0.06~\textrm{(stat.)}\pm0.15\textrm{(syst.)}$ to Draco dsph. Despite a relatively small number of red giant branch stars near the TRGB, the authors utilized more than a minimum of 100 stars - a criterion suggested by \citet{madore1995} for a robust determination of the TRGB magnitude.  Recently, \citet{madore2023} suggested that a population of $\sim100$ RGB stars is insufficient to provide reliable measurement of the TRGB magnitude. 

\begin{figure}
\centering
\includegraphics[width=1.0\columnwidth]{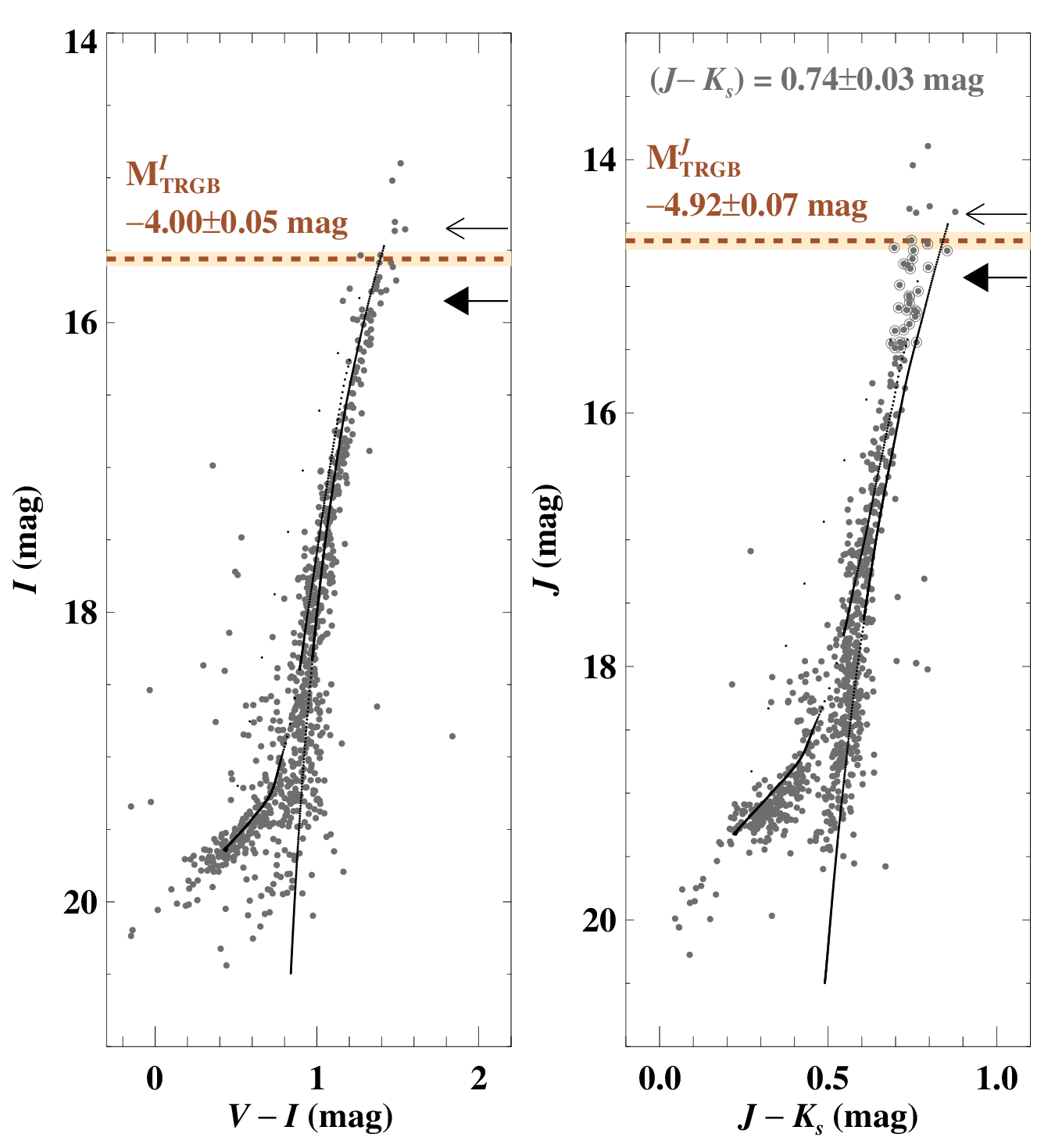}
\caption{Optical and NIR color--magnitude diagrams for the proper motion cleaned members of Draco dSph. {\it Left:} $V-I,I$ color--magnitude diagram. {\it Right:} $J-K_s,J$ color--magnitude diagram. The overplotted open circles represent red giants used to determine average color, $J-K_s=0.74\pm0.03$~mag, for TRGB measurement. The dashed lines represent the location of TRGB in Draco dSph in the optical and NIR and the shaded regions display their $\pm1\sigma$ uncertainties. The dotted lines show BaSTI stellar isochrones \citep{pietrinferni2021} with 10 Gyr age and metallicity of Draco dSph. The open and filled arrows display the expected location of TRGB corresponding to the distance modulus of 19.35 and 19.85 mag, respectively. These are the upper and lower limit of the distance moduli for Draco in the recent literature studies (see text).} 
\label{fig:trgb}
\end{figure}

We derived optical and NIR color--magnitude diagrams to determine an independent TRGB-based distance to Draco dSph. Although, \citet{kinemuchi2008} provided optical $VI$-band photometry for variables in Draco, we used homogeneous {\it Gaia} photometry of $\sim7000$ sources within $30\arcmin$ around the center of Draco for this analysis. The $G$-band magnitudes and (BP$-$RP) colors were used in the {\it Gaia} to Johnson-Cousin photometric transformations from \citet{pancino2022} to derive $V$ and $I$ band magnitudes. We apply proper motion cleaning as previously discussed in Section~\ref{subsec:cmd} for the NIR data. The extinction-correction in the $V$ and $I$ bands were estimated using $R_{V/I}=3.23/1.97 E(B-V)$ to be consistent with the coefficients adopted in the NIR bands. The optical and NIR color--magnitude diagrams are shown in Fig.~\ref{fig:trgb}. It is clearly evident that the number of RGB stars one-magnitude below the TRGB is very small, preventing an accurate and precise measurement of TRGB magnitude. 

Nevertheless, the color--magnitude diagrams shown in Fig.~\ref{fig:trgb} provide an opportunity to independently verify the accuracy of RRL-based distance to Draco dSph derived in this work. In the $I$-band, we adopted an absolute TRGB magnitude of $-4.00\pm0.05$ mag, which covers the range of most recent calibrations in the literature \citep[e.g.,][and references therein]{freedman2020, li2023, dixon2023}. In the NIR bands, the $J$-band TRGB magnitude was derived using $M_J=5.14-0.85[(J-K_s)-1.0]$ from \citet{freedman2020}. We used an average color $(J-K_s)=0.74\pm0.03$, using red giant stars in a magnitude-bin below the approximate location of the TRGB in Fig.~\ref{fig:trgb}, and found a value of $M^J_\textrm{TRGB}=-4.92\pm0.07$~mag. The apparent TRGB magnitude in $I$ and $J$ bands were derived using the above-mentioned absolute magnitudes and the RRL-based distance modulus to Draco dSph. In the color--magnitude diagrams, the discontinuity at the TRGB is clearly visible and is in excellent agreement with the derived apparent TRGB magnitudes in the optical and NIR bands, as shown with dashed lines in Fig.~\ref{fig:trgb}.
Adopting a significantly larger or smaller distance to Draco dSph results in an offset from the apparent discontinuity in the color--magnitude diagram. We also overplot the stellar isochrones from BaSTI{\footnote{\url{http://basti-iac.oa-abruzzo.inaf.it/index.html}}} (Bag of Stellar Tracks and Isochrones) for a constant age of 10 Gyr and metallicity of $Z=0.0004$ ([Fe/H]$=-1.90$~dex), and $Y=0.245$ using models with $\alpha$-enhanced calculations from \citet{pietrinferni2021}. The stellar isochrones were offset with the RRL distance and show a reasonable agreement with the apparent location of the TRGB in Draco dSph. This further confirms the accuracy of the distance derived in this work and the consistency among population II distance indicators in Draco dSph galaxy.

\section{Summary}
\label{sec:discuss}

We presented NIR time-series observations of pulsating variable stars in Draco dSph galaxy for the first-time. These observations were obtained using the WIRCam instrument at the 3.6-m Canada--France--Hawaii Telescope. Homogeneous NIR light curves of 212 RRL and 5 ACEP variables were fitted with templates resulting in accurate mean magnitudes in $JHK_s$ bands. We derived NIR PL relations for different subclasses of RRL stars and fundamental mode ACEP variables for the first-time in Draco dSph. The empirical PL relations for RRL stars exhibit a small scatter of $\sim0.05$~mag in $JHK_s$ bands that is comparable to those seen for RRL PL relations in globular clusters. 
The low dispersion of PL relation in NIR bands suggests at most about 0.2 dex spread in the metallicity of RRL stars in Draco dSph. We adopted a mean metallicity of $-1.98\pm0.10$~dex \citep{kirby2013} and calibrated NIR PLZ relations from \citet{bhardwaj2023} to determine distance modulus of $19.56\pm0.03$~mag to Draco dSph. The final distance to Draco $D_\textrm{Draco} = 81.55 \pm 0.98 \textrm{(statistical)} \pm 1.17 \textrm{(systematic)}$~kpc, is $1.5\%$ precise and the most accurate distance to this dSph galaxy. Using this RRL distance, the absolute magnitudes of the TRGB in $I$ and $J$ bands in the optical and NIR color--magnitude diagrams, respectively, agree well with their recent calibrations in the literature. However, a low number of red giant stars within the one magnitude of the TRGB prevents an accurate and precise determination of its magnitude, and in turn, an independent distance to Draco dSph. This paper is the first in a series that will provide time-series observations of pulsating variable stars in nearby galaxies. Precise RRL based distances to several nearby dwarf galaxies will allow an independent calibration of the TRGB, and a crucial test for the population II distance scale.  

\acknowledgements
%We thank the anonymous referee for useful comments that helped improve the manuscript.
We thank Dr. Tatiana Muraveva for kindly providing the variable star list in Draco dSph galaxy. This project has received funding from the European Union’s Horizon 2020 research and innovation programme under the Marie Skłodowska-Curie grant agreement No. 886298. This research was supported by the Munich Institute for Astro-, Particle and BioPhysics (MIAPbP) which is funded by the Deutsche Forschungsgemeinschaft (DFG, German Research Foundation) under Germany´s Excellence Strategy – EXC-2094 – 390783311.
CCN thanks the funding from the National Science and Technology Council (NSTC; Taiwan) under the contract 109-2112-M-008-014-MY3 and 112-2112-M-008-042.
Access to the CFHT was made possible by the Institute of Astronomy and Astrophysics, Academia Sinica.

%Based on observations obtained with WIRCam, a joint project of CFHT, Taiwan, Korea, Canada, France, at the Canada-France-Hawaii Telescope (CFHT) which is operated from the summit of Maunakea by the National Research Council of Canada, the Institut National des Sciences de l'Univers of the Centre National de la Recherche Scientifique of France, and the University of Hawaii. The observations at the Canada-France-Hawaii Telescope were performed with care and respect from the summit of Maunakea which is a significant cultural and historic site.
%HPS acknowledges grant 03(1428)/18/EMR-II from Council of Scientific and Industrial Research (CSIR), India.

\facility{CFHT-WIRCAM}
\software{\texttt{The IDL Astronomy User's Library} \citep{landsman1993},
\texttt{Astropy} \citep{astropy2013, astropy2018, astropy2022}, Libre-ESpRIT \citep{danoti1997}}

\vspace{100pt}

\bibliographystyle{aasjournal}
\bibliography{mybib_final.bib}

\end{document}